\documentclass[11pt,superscriptaddress,aps,prd,preprint]{revtex4}

\everymath{\displaystyle}
\usepackage{graphicx}

\usepackage[T1]{fontenc}
\usepackage{amsmath}
\usepackage{amssymb}
\usepackage{graphicx}

\newcommand{\bea}{\begin{eqnarray}}
\newcommand{\eea}{\end{eqnarray}}

\begin{document}

\title{On Gravitational Casimir Effect and Stefan-Boltzmann Law at Finite Temperature}

\author{S. C. Ulhoa}
\email{sc.ulhoa@gmail.com}
\affiliation{International Center of Physics, Instituto de F\'isica, Universidade de Bras\'ilia, 70910-900, Bras\'ilia, DF, Brazil}

\author{A. F. Santos}
\email{alesandroferreira@fisica.ufmt.br}
\affiliation{Instituto de F\'{\i}sica, Universidade Federal de Mato Grosso,\\
78060-900, Cuiab\'{a}, Mato Grosso, Brazil}

\author{T. F. Furtado}
\email{tarciofla@hotmail.com}
\affiliation{Instituto de F\'isica, Universidade de Bras\'ilia, 70910-900, Bras\'ilia, DF, Brazil}

\author{F. C. Khanna \footnote{Professor Emeritus - Physics Department, Theoretical Physics Institute, University of Alberta\\
Edmonton, Alberta, Canada}}
\email{khannaf@uvic.ca}
\affiliation{Department of Physics and Astronomy, University of
Victoria,BC V8P 5C2, Canada}

\begin{abstract}

Gravitons are described by the propagator in Teleparallel gravity in nearly flat space-time. Finite temperature is introduced by using Thermo Field Dynamics formalism. The Gravitational Casimir effect and Stefan-Boltzmann law are calculated as a function of temperature. Then an equation of state for gravitons is determined.

\end{abstract}

\maketitle

\section{Introduction} \label{sec.1}

The teleparallell theory of gravity is similar to the general theory of gravity due to Einstein \cite{HS,HS2,hehl,blag} except that the it has a well defined expression of energy-momentum tensor \cite{maluf0}. In addition the Teleparallel gravity has a successful quantum approach using Weyl prescription \cite{ulhoa}. Using a path integral approach a different propagator in teleparallel gravity is obtained \cite{ulhoa1}. Furthermore it is possible to introduce finite temperature effects in this formalism.

There are several approaches to introduce finite temperature formulation in a theory that can be quantized. Such a method involves doubling the physical space, Thermo Field Dynamics (TFD) \cite{Umezawa1, Umezawa2, Umezawa22, Khanna1, Khanna2, Santana1, Santana2} . This thermal theory is obtained by using Bogoliubov transformation and it is a real time formulation. The thermal state is obtained by doubling of Hilbert space, i.e., real space and a tilde space. The tilde space introduces temperature in the problem.

The objective of the paper is to use teleparallel theory of gravity and formulation of TFD to calculate the gravitational Casimir effect and Stefan-Boltzmann Law. Originally the Casimir effect was observed by H. Casimir for parallel plates \cite{Casimir} which are attracted due to vacuum fluctuations of the electromagnetic field. Subsequent experiments have established this effect to a high degree of accuracy \cite{Sparnaay, Lamoreaux, Mohideen}. Experiments have observed both Casimir effect and Stefan-Boltzmann law experimentally for standard field theories. Gravitational field effects have been considered in gravitational field \cite{Quach, gem,Elizalde,Elizalde1}, violation of Lorentz theory \cite{LV1, LV2, LV3} and in Kalb-Ramond field \cite{Ademir}, among others.

This paper is organized as follows. In section II, details of the teleparallel gravity are described. In section III, details of the TFD formalism are defined in detail. In section IV, the Green function is defined using the graviton propagator for the teleparallel gravity. In section V, the Casimir effect and the Stefan-Boltzmann law are obtained using finite temperature procedure. Then the first law of thermodynamics for gravitons is obtained. Finally in section VI, some concluding remarks are given.

\section{Teleparallel Gravity} \label{sec.2}

Teleparallel gravity has unique features such as a well defined expression for the gravitational energy-momentum tensor. This alternative theory is constructed in the framework of Weitzenb\"{o}ck geometry that is described by torsion and a vanishing curvature. Thus a Weitzenb\"ockian manifold is endowed with a Cartan connection \cite{cartan}, defined by
\bea
\Gamma_{\mu\lambda\nu}=e^{a}\,_{\mu}\partial_{\lambda}e_{a\nu},
\eea
where $e^{a}\,_{\mu}$ is the tetrad field. Such a variable assures two fundamental symmetries in teleparallel gravity, Lorentz symmetry denoted by latin indices and diffeomorphism represented by greek ones.

It is important to note that the Cartan connection is curvature free but it has the torsion tensor
\begin{equation}
T^{a}\,_{\lambda\nu}=\partial_{\lambda} e^{a}\,_{\nu}-\partial_{\nu}
e^{a}\,_{\lambda}\,. \label{4}
\end{equation}
On the other hand, Weitzenb\"ock geometry is intrinsically related to Riemann geometry by the following mathematical identity
\begin{equation}
\Gamma_{\mu \lambda\nu}= {}^0\Gamma_{\mu \lambda\nu}+ K_{\mu
\lambda\nu}\,, \label{2}
\end{equation}
where ${}^0\Gamma_{\mu \lambda\nu}$ are the Christoffel symbols and $K_{\mu \lambda\nu}$ is the contortion tensor given by
\begin{eqnarray}
K_{\mu\lambda\nu}&=&\frac{1}{2}(T_{\lambda\mu\nu}+T_{\nu\lambda\mu}+T_{\mu\lambda\nu})\,,\label{3}
\end{eqnarray}
with $T_{\mu \lambda\nu}=e_{a\mu}T^{a}\,_{\lambda\nu}$. Such an identity leads to the following relation
\begin{equation}
eR(e)\equiv -e({1\over 4}T^{abc}T_{abc}+{1\over
2}T^{abc}T_{bac}-T^aT_a)+2\partial_\mu(eT^\mu)\,,\label{eq5}
\end{equation}
where $e$ is the determinant of the tetrad field and $R$ is the scalar mode of Christoffel symbols. Then a gravitational theory, dynamically equivalent to general relativity, is established by the following Lagrangian density
\begin{equation}
\mathfrak{L}(e_{a\mu})=-\kappa\,e \Sigma^{abc}T_{abc} -\mathfrak{L}_M\;, \label{6}
\end{equation}
where $\kappa=1/(16 \pi)$, $\mathfrak{L}_M$ is the Lagrangian
density of matter fields and $\Sigma^{abc}$ is given by
\begin{equation}
\Sigma^{abc}={1\over 4} (T^{abc}+T^{bac}-T^{cab}) +{1\over 2}(
\eta^{ac}T^b-\eta^{ab}T^c)\;, \label{7}
\end{equation}
with $T^a=e^a\,_\mu T^\mu$. It is worth to point out that in the Weitzenb\"och geometry it is possible to obtain a large number of invariants compared to the Riemannian case.

If a derivative of the Lagrangian density with respect to the tetrad field is performed, it yields
\begin{equation}
\partial_\nu\left(e\Sigma^{a\lambda\nu}\right)={1\over {4\kappa}}
e\, e^a\,_\mu( t^{\lambda \mu} + T^{\lambda \mu})\;, \label{10}
\end{equation}
with $e\,e^a\,_\mu T^{\lambda \mu}=\frac{\partial \mathfrak{L}_M}{\partial e_{a\lambda}}$ and
\begin{equation}
t^{\lambda \mu}=\kappa\left[4\,\Sigma^{bc\lambda}T_{bc}\,^\mu- g^{\lambda
\mu}\, \Sigma^{abc}T_{abc}\right]\,. \label{11}
\end{equation}
This is the gravitational energy-momentum tensor \cite{maluf,jose}. It should be noted that the quantity $\Sigma^{a\lambda\nu}$ is skew-symmetric in the last two indices, i. e.,
$\partial_\lambda\partial_\nu\left(e\Sigma^{a\lambda\nu}\right)\equiv0\,$ which implies a conservation law that leads to the energy-momentum vector 
\begin{equation}
P^a = \int_V d^3x \,e\,e^a\,_\mu(t^{0\mu}+ T^{0\mu})\,, \label{14}
\end{equation}
that is a well defined expression for energy and momentum. It is a vector under Lorentz transformations.

\section{Thermo Field Dynamics (TFD)} \label{sec.3}

This section includes brief details of Thermo Field Dynamics (TFD) \cite{Umezawa1, Umezawa2, Umezawa22, Khanna1, Khanna2, Santana1, Santana2}. TFD is a real time formalism of Quantum Field Theory at finite temperature. The thermal average of an observable is given in an extended Hilbert space. There are two necessary ingredients of TFD: the doubling of the degrees of freedom in Hilbert space and the Bogoliubov transformations. The doubling is defined by the tilde ($\sim$) conjugation. Each operator in the Hilbert space $S$ has a fictitious Hilbert space $\tilde{S}$. The Bogoliubov transformation combines the two spaces.

The creation ($a^\dagger$) and annihilation operators ($a$) in the standard Hilbert space and tilde operators, $\tilde a^\dagger$ and $\tilde{a}$ are combined in the extended space using Bogoliubov transformations. For bosons these are 
\bea
a(k, \alpha)&=&u(\alpha)a(k)-v(\alpha)\tilde a^\dagger(k),\\
\tilde a^\dagger(k, \alpha)&=&u(\alpha)\tilde a^\dagger(k)-v(\alpha) a(k),
\eea
and the algebraic rules for the operators are 
\bea
\left[a(k, \alpha), a^\dagger(p, \alpha)\right]=\delta^3(k-p),\quad\quad\quad \left[\tilde{a}(k, \alpha), \tilde{a}^\dagger(p, \alpha)\right]=\delta^3(k-p),\label{ComB}
\eea
and other commutation relations are null. The algebraic rules for these thermal operators are the same as obtained in the quantum field theory at zero temperature.  The quantities $u(\alpha)$ and $v(\alpha)$ are related to the Bose distribution given as
\bea
v^2(\alpha)=\frac{1}{e^{\alpha\omega}-1}, \quad\quad u^2(\alpha)=1+v^2(\alpha).\label{phdef}
\eea
Here $\omega=\omega(k)$ and $\alpha=1/\beta$ with $\beta=1/(k_B T)$ and $k_B$ is the Boltzmann constant and $T$ is the temperature. There is a similar Bogoliubov transformation for fermions.

A doublet notation is introduced by 
\bea
\left( \begin{array}{cc} a(k, \alpha)  \\ \tilde a^\dagger(k, \alpha) \end{array} \right)={\cal B}(\alpha)\left( \begin{array}{cc} a(k)  \\ \tilde a^\dagger(k) \end{array} \right),
\eea
where ${\cal B}(\alpha)$ is the Bogoliubov transformation given as
\bea
{\cal B}(\alpha)=\left( \begin{array}{cc} u(\alpha) & -v(\alpha) \\
-v(\alpha) & u(\alpha) \end{array} \right),
\eea
The $\alpha$ parameter is assumed as the compactification parameter defined by $\alpha=(\alpha_0,\alpha_1,\cdots\alpha_{D-1})$. The effect of temperature is described by the choice $\alpha_0\equiv\beta$ and $\alpha_1,\cdots\alpha_{D-1}=0$, where $D$ are the space-time dimensions. Any propagator in the TFD formalism may be written in terms of the $\alpha $-parameter. For the scalar field propagator, as an example, the Green function is
\bea
G_0^{(AB)}(x-x';\alpha)=i\langle 0,\tilde{0}| \tau[\phi^A(x;\alpha)\phi^B(x';\alpha)]| 0,\tilde{0}\rangle,
\eea
where $A, B=1, 2$ are the indices that define the double space and
\bea
\phi(x;\alpha)&=&{\cal B}(\alpha)\phi(x){\cal B}^{-1}(\alpha).
\eea
 In the thermal vacuum $|0(\alpha)\rangle={\cal B}(\alpha)|0,\tilde{0}\rangle$, the propagator becomes
\bea
G_0^{(AB)}(x-x';\alpha)&=&i\langle 0(\alpha)| \tau[\phi^A(x)\phi^B(x')]| 0(\alpha)\rangle,\nonumber\\
&=&i\int \frac{d^4k}{(2\pi)^4}e^{-ik(x-x')}G_0^{(AB)}(k;\alpha),
\eea
where
\bea
G_0^{(AB)}(k;\alpha)={\cal B}^{-1}(\alpha)G_0^{(AB)}(k){\cal B}(\alpha),
\eea
with
\bea
G_0^{(AB)}(k)=\left( \begin{array}{cc} G_0(k) & 0 \\
0 & \xi G^*_0(k) \end{array} \right),
\eea
and $G_0(k)=\frac{1}{k^2}$. Here $\xi = -1$ for bosons and $\xi = +1$ for fermions. The non-tilde variables describe the physical quantities. Then
\bea
G_0^{(11)}(k;\alpha)=G_0(k)+\xi u^2(k;\alpha)[G^*_0(k)-G_0(k)],
\eea
where $u^2(k;\alpha)$, the generalized Bogoliubov transformation \cite{GBT}, is
\bea
u^2(k;\alpha)=\sum_{s=1}^d\sum_{\lbrace\sigma_s\rbrace}2^{s-1}\sum_{l_{\sigma_1},...,l_{\sigma_s}=1}^\infty(-\eta)^{s+\sum_{r=1}^sl_{\sigma_r}}\,\exp\left[{-\sum_{j=1}^s\alpha_{\sigma_j} l_{\sigma_j} k^{\sigma_j}}\right],\label{BT}
\eea
with $d$ being the number of compactified dimensions, $\eta=1(-1)$ for fermions (bosons), $\lbrace\sigma_s\rbrace$ denotes the set of all combinations with $s$ elements and $k$ is the 4-momentum.

\section{Gravitational Casimir Effect and Stefan-Boltzmann Law} \label{sec.4}

In this section the Stefan-Boltzmann law and the Casimir effect at zero and finite temperature are calculated in the teleparallel gravity framework. The free Lagrangian of the teleparallel gravity is 
\begin{equation}
\mathfrak{L}_g=-ke\Sigma^{abc}T_{abc}\,,
\end{equation}
and using the weak field approximation, i.e.,
\begin{equation}
g_{\mu\nu}=\eta_{\mu\nu}+h_{\mu\nu},
\end{equation}
the graviton propagator is obtained as
\begin{equation}
\langle e_{b\lambda}, e_{d\gamma} \rangle=\Delta_{bd\lambda\gamma} = \frac{i\eta_{bd}}{\kappa q^{\lambda} q^{\gamma}}.
\end{equation}
Some details are given in \cite{ulhoa1}. This leads to the Green function
\bea
G_0(x,x')=-i\Delta_{bd\lambda\gamma}\,g^{\lambda\gamma}\eta^{bd},
\eea
which explicitly is
\begin{equation}
G_0(x,x')= -\frac{i64\pi}{q^{2}}\,,
\end{equation}
with $q=x-x'$. It is worth comparing the graviton propagator in TEGR and general relativity. There is remarkable difference between the graviton propagator in general relativity that has the form $$\Delta^{\mu\nu\lambda\gamma}=\frac{i}{2}\left(\eta^{\mu\lambda} \eta^{\nu\gamma}+ \eta^{\mu\gamma} \eta^{\nu\lambda}-\eta^{\mu\nu} \eta^{\lambda\gamma}\right)G_0(q)\,,$$ where $G_0(q)$ is the Green function of a massless scalar field. Such a difference would be expected since the symmetry in the two theories are not the same.

Now it is necessary to calculate the mean value of $t^{\lambda \mu}$ (the gravitational energy-momentum tensor). In the weak field approximation, expanding expression (\ref{11}) leads to
\begin{eqnarray}
	T^{bc\lambda}T_{bc}\,^{\mu} &=& g^{\mu \alpha}(\partial^{\gamma}e^{b\lambda} - \partial^{\lambda}e^{b\gamma})(\partial_{\gamma}e_{b\alpha}-\partial_{\alpha}e_{b\gamma})\,,\nonumber\\
	T^{bc\lambda}T_{bc}\,^{\mu}g_{\mu\lambda} &=& 2\partial^{\gamma}e^{b\alpha}(\partial_{\gamma}e_{b\alpha}-\partial_{\alpha}e_{b\gamma} )\,,\nonumber \\
    	T^{bc\lambda}T_{bc}\,^{\mu}g_{\mu\lambda} &=& 2\partial^{\gamma}e^{b\alpha}(\partial_{\gamma}e_{b\alpha}-\partial_{\alpha}e_{b\gamma} ).
\end{eqnarray}
These quantities are second order in the tetrad field while the terms
\begin{eqnarray}
	T^{cb\lambda}T_{bc}\,^{\,\,\,\mu} &=& e^{b}_{\,\,\,\gamma}\, e_{c}^{\,\,\,\alpha}(\partial^{\gamma}e^{c\lambda} - \partial^{\lambda}e^{c\gamma})(\partial_{\alpha}e_{b}^{\,\,\,\mu}-\partial^{\mu}e_{b\alpha})\,, \nonumber\\
	T^{\lambda bc}T_{bc}\,^{\,\,\,\mu} &=& e_{a}^{\,\,\,\lambda}e^{b}_{\,\,\,\alpha}(\partial^{\alpha}e^{a\gamma} -\partial^{\gamma}e^{a\alpha})(\partial_{\gamma}e_{b}^{\,\,\,\mu}-\partial^{\mu}e_{b\gamma}) \,, \nonumber\\
	e^{b\lambda}\,T^{c}\,T_{bc}\,^{\,\,\,\mu} &=& e^{b\lambda}e_{a\alpha}(\partial^{\alpha}e^{a\gamma} -\partial^{\gamma}e^{a\alpha})(\partial_{\gamma}e_{b}^{\,\,\,\mu}-\partial^{\mu}e_{b\gamma})\,,\nonumber\\
	\eta^{bc}T^{\gamma}T_{bc}\,^{\,\,\,\mu} &=& \eta^{bc}e_{a\alpha}e_{c\gamma}(\partial^{\alpha}e^{a\lambda} -\partial^{\lambda}e^{a\alpha})(\partial^{\gamma}e_{b}^{\,\,\,\mu}-\partial^{\mu}e_{b}^{\,\,\,\gamma})\,,
\end{eqnarray}
are of higher order in the field. These are dropped in the weak field approximation. Then the gravitational energy-momentum tensor is
\begin{eqnarray}
t^{\lambda\mu} &=& \kappa\Bigl[g^{\mu\alpha}\partial^{\gamma}e^{b\lambda}\partial_{\gamma}e_{b\alpha} - g^{\mu\lambda}\partial^{\alpha}e^{b\lambda}\partial_{\gamma}e_{b\alpha} - g^{\mu\alpha}(\partial^{\lambda}e^{b\gamma}\partial_{\gamma}e_{b\alpha} - \partial^{\lambda}e^{b\gamma}\partial_{\alpha}e_{b\gamma})\nonumber\\
        & &-2g^{\lambda\mu}\partial^{\gamma}e^{b\alpha}(\partial_{\gamma}e_{b\alpha}-\partial_{\alpha}e_{b\gamma})\Bigl]\,.
\end{eqnarray}
To avoid a product of field operators at the same space-time point, the expectation value of energy-momentum tensor is
\bea
\langle t^{\lambda\mu}(x)\rangle&=& \langle 0|t^{\lambda\mu}(x)|0\rangle,\nonumber\\
&=& \lim_{x^\mu\rightarrow x'^\mu} 4i\kappa\left(-5g^{\lambda\mu}\partial'^{\gamma}\partial_{\gamma} +2g^{\mu\alpha}\partial'^{\lambda}\partial_{\alpha}\right)G_{0}(x-x')\,,
\eea
where $\langle e_{c}^{\,\,\,\lambda}(x), e_{b\alpha}(x') \rangle = i\eta_{cb}\,\delta^{\lambda}_{\alpha}\,G_{0}(x-x')$, with $G_{0}(x-x')$ being the Green function.

In the  TFD formalism, using the tilde conjugation rules, the vacuum average of the gravitational energy-momentum tensor is
\bea
\langle t^{\lambda\mu(AB)}(x;\alpha)\rangle=\lim_{x\rightarrow x'} 4i\kappa\left(-5g^{\lambda\mu}\partial'^{\gamma}\partial_{\gamma} +2g^{\mu\alpha}\partial'^{\lambda}\partial_{\alpha}\right)G_{0}^{(AB)}(x-x';\alpha).
\eea
It is impossible to get an analogues expression  for general relativity since there is no energy-momentum tensor. However there is a propagator for the graviton. It is shared by other gravitational theories where the concept of gravitational energy is well established. Gravitoelectromagnetism (GEM) is certainly one theory where this is possible.

Following the Casimir prescription, the physical energy-momentum tensor is given by
\bea
{\cal T}^{\lambda\mu (AB)}(x;\alpha)=\langle t^{\lambda\mu(AB)}(x;\alpha)\rangle-\langle t^{\lambda\mu(AB)}(x)\rangle.
\eea
Explicitly
\bea
{\cal T}^{\lambda\mu (AB)}(x;\alpha)=\lim_{x\rightarrow x'} 4i\kappa\left(-5g^{\lambda\mu}\partial'^{\gamma}\partial_{\gamma} +2g^{\mu\alpha}\partial'^{\lambda}\partial_{\alpha}\right)\overline{G}_{0}^{(AB)}(x-x';\alpha),
\eea
where 
\bea
\overline{G}_0^{(AB)}(x-x';\alpha)=G_0^{(AB)}(x-x';\alpha)-G_0^{(AB)}(x-x').
\eea
In the Fourier representation 
\bea
\overline{G}_0^{(AB)}(x-x';\alpha)=\int \frac{d^4k}{(2\pi)^4}e^{-ik(x-x')}\overline{G}_0^{(AB)}(k;\alpha),
\eea
such that its physical component is
\bea
\overline{G}_0^{(11)}(k;\alpha)=u^2(\alpha)\left[G_0(k)-G_0^*(k)\right],
\eea
where $u^2(\alpha)$ is the generalized Bogoliubov transformation given in eq. (\ref{BT}).

Now some applications are considered for different choices of the $\alpha$-parameter.

\subsection{Gravitational Stefan-Boltzmann law}

A first application is the case $\alpha=(\beta,0,0,0)$, such that the Bogoliubov transformation is given as
\bea
u^2(\beta)=\sum_{j_0=1}^\infty e^{-\beta k^0 j_0}.
\eea
Then the physical component of the Green function is
\bea
\overline{G}_0^{(11)}(x-x';\beta)&=&\int \frac{d^4k}{(2\pi)^4}e^{-ik(x-x')}\sum_{j_0=1}^\infty e^{-\beta k^0 j_0}\left[G_0(k)-G_0^*(k)\right],\nonumber\\
&=&2\sum_{j_0=1}^\infty G_0^{(11)}\left(x- x'-i\beta j_0 n_0\right),\label{1GF}
\eea
where $n_0^\mu=(1,0,0,0)$. The vacuum expectation value of the gravitational energy-momentum tensor at finite temperature is
\begin{equation}
{\cal T}^{\lambda\mu (11)}(x;\beta)=\lim_{x\rightarrow x'}\sum_{j_0=1}^{\infty} 8i\kappa(-5g^{\lambda\mu}\partial'^{\gamma}\partial_{\gamma} +2g^{\mu\alpha}\partial'^{\lambda}\partial_{\alpha})\, G_0^{(11)}\left(x- x'-i\beta j_0 n_0\right)\,,
\end{equation}	
At this point a nearly flat space-time, that represents gravitons precisely, may be analyzed. With $\lambda=\mu=0$ and using the Riemann Zeta function~\cite{Elizalde2, Elizalde3}
\bea
\zeta(4)=\sum_{j_0=1}^\infty\frac{1}{j_0^4}=\frac{\pi^4}{90},\label{zetaf}
\eea
the  gravitational Stefan-Boltzmann law is given by
\begin{equation}
 E(T)=\dfrac{64}{15}\pi^{4}T^{4}\,,
\end{equation}
where $E(T)\equiv {\cal T}^{00(11)}(x;\alpha) $. The first law of Thermodynamics is formulated as
\begin{equation}
    E+P = T\left( \dfrac{\partial P}{\partial T} \right)_{V}\,,
\end{equation}
then
\begin{equation}
   \sigma T^{4} + P = T\left( \dfrac{\partial P}{\partial T} \right)_{V}
\end{equation}
where $\sigma = \dfrac{64\pi^{4}}{15}$. As a consequence, $P = \dfrac{\sigma T^{4}}{3}$ which leads to the equation of state
\begin{equation}
    P = \dfrac{E}{3} \,.
\end{equation}
The equation of state for both gravitons and photons is the same.

\subsection{Gravitational Casimir effect at zero temperature}

In the framework of TFD formalism the Casimir effect at zero temperature is determined when $\alpha=(0,0,0,i2d)$. Then the Bogoliubov transformation becomes
\bea
u^2(d)=\sum_{l_3=1}^\infty e^{-i2dk^3l_3}
\eea
and the Green function is given as
\bea
\overline{G}_0^{(11)}(x-x';d)&=&2\sum_{l_3=1}^\infty G_0^{(11)}\left(x-x'-2dl_3z\right).\label{2GF}
\eea
Then the energy-momentum tensor is
\bea
{\cal T}^{\lambda\mu (11)}(x;d)=\lim_{x\rightarrow x'}\sum_{l_3=1}^{\infty} 8i\kappa(-5g^{\lambda\mu}\partial'^{\gamma}\partial_{\gamma} +2g^{\mu\alpha}\partial'^{\lambda}\partial_{\alpha})\, G_0^{(11)} \left(x- x'-2dl_3 n_3\right),
\eea
where $n_3=(0,0,0,1)$. Then for $\lambda=\mu=0$ the gravitational Casimir energy (at zero temperature) associated with the gravitons is
\begin{equation}
E_{c}(d)=-\dfrac{4\pi^{4}}{45d^{4}}\,,
\end{equation}
where the Riemann Zeta function has been used and $E_{c}(d) \equiv {\cal T}^{00 (11)}(x;d)$. Under such conditions and with $\lambda=\mu=3$ the gravitational Casimir pressure is calculated as well, it reads
\begin{equation}
 P_{c}(d)=-\dfrac{4\pi^{4}}{15d^{4}}\,,
\end{equation}
where $P_{c}(d) \equiv {\cal T}^{33 (11)}(x;d)$. Here the Casimir force between the plates is negative, then it is an attractive force, similar to the case of the electromagnetic field.

\subsection{Gravitational Casimir effect at finite temperature}

The effect of temperature in the Casimir effect is introduced by taking $\alpha=(\beta, 0, 0,i2d)$. Then 
\bea
v^2(k^0,k^3;\beta,d)&=&v^2(k^0;\beta)+v^2(k^3;d)+2v^2(k^0;\beta)v^2(k^3;d),\nonumber\\
&=&\sum_{j_0=1}^\infty e^{-\beta k^0j_0}+\sum_{l_3=1}^\infty e^{-i2dk^3l_3}+2\sum_{j_0,l_3=1}^\infty e^{-\beta k^0j_0-i2dk^3l_3}\label{eq51}
\eea
and
\bea
\overline{G}_0^{(11)}(x-x';\beta,d)&=&4\sum_{j_0,l_3=1}^\infty G_0^{(11)}\left(x-x'-i\beta j_0n-2dl_3z\right)\label{3GF}
\eea
are the Bogoliubov transformation and the Green function, respectively. The first term in eq. (\ref{eq51}) leads to the Stefan-Boltzmann law, the second term to the Casimir effect at zero temperature and the third term to the Casimir effect at finite temperature. Considering only the third term, the energy-momentum tensor becomes
\bea
{\cal T}^{\lambda\mu (11)}(\beta;d)&=&\lim_{x\rightarrow x'}\sum_{j_0,l_3=1}^\infty 16i\kappa(-5g^{\lambda\mu}\partial'^{\gamma}\partial_{\gamma} +2g^{\mu\alpha}\partial'^{\lambda}\partial_{\alpha})\nonumber\\
&\times&G_0^{(11)}\left(x-x'-i\beta j_0n-2dl_3z\right).
\eea
 The Casimir energy at finite temperature for gravitons is
\begin{equation}
 E_{c}(\beta, d)={\cal T}^{00 (11)}(\beta;d)=-256\sum_{j_0, l_3}\dfrac{[(2d l_3)^{2}-3(\beta j_0)^{2}]}{[(2d l_3)^{2}+(\beta j_0)^{2}]^{3}}\,,
\end{equation}
and the Casimir pressure at finite temperature is
\begin{equation}
    P_{c}(\beta, d)={\cal T}^{33 (11)}(\beta;d)=-256\sum_{j_0, l_3}\dfrac{[3(2d l_3)^{2}-(\beta j_0)^{2}]}{[(2d l_3)^{2}+(\beta j_0)^{2}]^{3}}\,,
\end{equation}
It is to be noted that these results were obtained using a metric tensor of an approximate Minkowski space-time. In addition $E_{c}(\beta, d)$ recovers the dependency of $T^4$ for $d\rightarrow 0$, similarly it tends to be proportional to $d^{-4}$ when $T\rightarrow 0$. The same behavior is observed for $P_{c}(\beta, d) $.

\section{Conclusion} \label{sec.5}

The graviton propagator is calculated using teleparallel gravity and this leads to a Green function. Details of TFD formalism are given to introduce temperature. Stefan-Boltzmann law and Casimir effect are calculated at finite temperature. This leads to finding pressure as a function of temperature. These results are obtained for a nearly flat space-time. These results play an important role in comparing with experimental results obtained for systems in outer space. An extension of this program for different space-time points will be carried out later. In addition the first law of thermodynamics is used to establish the dependency of the gravitational pressure on the temperature. The equation of state is found identical to that obtained for photons.

\section*{Acknowledgments}

This work by A. F. S. is supported by CNPq projects 308611/2017-9 and 430194/2018-8.

\end{document}